\pdfoutput=1
\documentclass[prd,preprint,onecolumn,nofootinbib,amsmath,amssymb]{revtex4}

\usepackage{amsmath,amssymb}
\usepackage{cancel}

\usepackage{hyperref}
\usepackage{natbib}
\usepackage{anysize}
\usepackage{graphicx}
\usepackage{slashed}
\usepackage[normalem]{ulem}

\marginsize{3cm}{3cm}{3cm}{3cm}

\begin{document}

\title{$Hb\bar{b}$ production in Composite Higgs Models}

\author{Mikael Chala}
\author{Jos\'e Santiago}
\affiliation{
CAFPE and Departamento de F\'{\i}sica Te\'orica y del Cosmos,
\\
Universidad de Granada, E-18071 Granada, Spain
}

\begin{abstract}
New vector-like quarks with electric charge $2/3$ and $-1/3$ can be singly produced at hadron colliders 
through the exchange of
a color octet vector resonance in models of strong electroweak
symmetry breaking. 
We 
show that electroweak symmetry breaking
effects can have a significant impact on the decay pattern of these new quarks. In particular, single production of charge $-1/3$
fermion resonances, mediated by a color octet vector resonance, typically results in an $H b\bar{b}$ final state
with a sizeable cross section and very distinctive kinematics. 
We consider the leading $H\to b\bar{b}$ decay and show that the $4b$
signal can be very efficiently disentangled from the
background: heavy octet masses of up to 3 TeV can be tested with the
data already collected at the LHC and up to 5 TeV with an integrated 
luminosity of
100 fb$^{-1}$ at $\sqrt{s}=14$ TeV. We also discuss 
the kinematical differences between the
$Hb\bar{b}$ production in models of strong electroweak symmetry
breaking and supersymmetric models and the implications on the
phenomenology of non-minimal composite Higgs models. 
\end{abstract}

\maketitle

\section{Introduction}

Composite Higgs models are among the leading candidates to dynamically
explain the origin of the electroweak symmetry breaking (EWSB) scale.
The absence of any significant departure from the Standard Model (SM)
predictions in current LHC searches, although somewhat disappointing,
was not unexpected~\cite{Santiago:2011dn}. 
The reason is that the constraints that electroweak precision tests typically
impose on the scale of the new resonances are stringent enough to make
their discovery in the low energy phase of the LHC highly
unlikely.
New vector resonances are expected to have masses in the
multi-TeV range, well above the current LHC
reach~\cite{Agashe:2004rs,Agashe:2005dk,Giudice:2007fh}.  
Naturalness~\cite{Matsedonskyi:2012ym,Redi:2012ha,
Marzocca:2012zn,Pomarol:2012qf,Panico:2012uw} 
and
compatibility with electroweak precision
tests~\cite{Carena:2006bn,Carena:2007ua,Anastasiou:2009rv} on the
other hand  
predict fermion resonances to be relatively light 
with typical masses below the TeV scale. 

These new fermion resonances, called top partners, are vector-like
quarks that mix strongly with the SM top quark. They are arranged in
multiplets of the unbroken global symmetries of the composite sector
which are likely to include at the very least an $SO(4)$ custodial
symmetry. Top partners can be efficiently searched for at the LHC
through their pair or single (electroweak)
production~\cite{Carena:2007tn,Contino:2008hi,
AguilarSaavedra:2009es,Mrazek:2009yu,DeSimone:2012fs,Vignaroli:2012nf} and
current data are already starting to probe part of the region
of parameter space allowed by indirect constraints.
It has been recently pointed out that top partners can be also
singly produced via the s-channel exchange of a color octet vector
resonance. This production mechanism can be competitive with the
previous ones and has the advantage of probing different aspects of
the composite sector~\cite{Barcelo:2011wu,Bini:2011zb,Carmona:2012my}.
Note that, even if \textit{a priori} the
composite sector does not need to have color octet vector resonances,
they naturally occur in models in which partial
compositeness~\cite{Kaplan:1991dc,Contino:2006nn} 
is realized.  

In this article we will show that if the bottom partners, the fermion
resonances responsible for the mass of the bottom quark, are not much
heavier than the top partners there can be a sizeable production of $H
b\bar{b}$ events in composite Higgs models~\cite{Chala:2013ula}. 
The large cross section
has its origin in the single production of a bottom partner, via the
s-channel exchange of a heavy color octet vector, followed by the
decay into a Higgs boson and the SM bottom quark:
\begin{equation}
pp\to G^\ast \to B_H\bar{b}+b\bar{B}_H \to H b \bar{b},
\end{equation}
where we have denoted by $G^\ast$ 
the vector resonance, called from now on heavy gluon, 
and $B_H$ the bottom partner (heavy bottom).
If flavor is realized through partial compositeness 
the bottom quark is lighter than the top
quark because it is less composite and not because its partners are
much heavier. 
EWSB effects, on the other hand, are quite relevant to correctly
describe the phenomenology of bottom partners. Due to the 
small mixing between the bottom quark and its partners, the presence
of other fermion resonances and the sizeable Yukawa couplings among
them can have a large impact in the decay pattern of the heavy
fermions. We explain the origin of this effect and its possible
relevance in models of strong EWSB in Appendix~\ref{appendix:A}. In
the example described in the appendix 
the only resonance accessible at the LHC is the
partner of the $b_R$ but it has a phenomenology wildly different from a
vector-like singlet. This  
shows that when large couplings are expected among the new
particles -like in models of strong EWSB- heavier states beyond the
LHC reach can have a huge impact on the phenomenology of the lighter
resonances that we can access experimentally. Thus we see that 
simplified models which
consider only the lightest resonances in the spectrum, although an
interesting first approach to models of new physics, can have a
phenomenology that grossly
deviates from the actual phenomenology of the full models.

$Hb\bar{b}$ production with $H\to b\bar{b}$ decay
has been proposed as a useful channel 
to search for neutral scalars in supersymmetric
models at large
$\tan \beta$~\cite{DiazCruz:1998qc,Dai:1994vu,Balazs:1998nt,Carena:1998gk}. 
The $\tan \beta$ enhancement of
the cross section is however dwarfed by the huge QCD background and
the difficulty of a clean trigger (imposing a hard cut on the $p_T$ of 
all four $b$ jets reduces the signal to negligible levels). In our case, the
large masses of the intermediate particles ($G^\ast$ and $B_H$) change the
picture completely. We can impose very stringent cuts on the $p_T$ of
the $b$ jets, which ensure a clean triggering and a very efficient
reduction of the
background. We will show that the specific kinematics of the process
allows for an excellent reconstruction of both the bottom partner and
the heavy gluon.
Other searches that are sensitive to the signature we study in this article,
again motivated by supersymmetric models, involve final states with a
large number of $b$-jets plus a sizeable amount of missing energy. 
Our signal does not have real missing energy but the large energy of
the quarks involved represent a non-negligible 
source of fake missing $E_T$. We will show that simple
modifications of current multi-$b$ searches, like the requirement of
harder $b$-jets and/or less missing energy, can turn these analyses
into very powerful probes of composite Higgs models.

The main results of this work are the expected $95\%$ C.L. exclusion
bounds on the single production cross section of $B_H$ (via a heavy
gluon) times its branching fraction into $Hb\bar{b}$ and the discovery
reach, that we report as a function of the main parameters of the
model. 
We have found that, assuming $M_B\approx M_{G^\ast}/2$, 
masses up to $\sim 3$ ($2.75$) TeV for the heavy gluon can be excluded
(discovered) with the data already collected at the LHC. This extends
up to $\sim 5$ and $4.5$ TeV of exclusion and discovery limits for the LHC
with $\sqrt{s}=14$ TeV and an integrated luminosity of 100
fb$^{-1}$.

Finally, we will argue that in composite
Higgs models with an extended scalar sector~\cite{Gripaios:2009pe,Mrazek:2011iu,Frigerio:2012uc,Bertuzzo:2012ya,Chala:2012af,Vecchi:2013bja}, a similar process
in which the Higgs boson is replaced by a mostly singlet composite
scalar might be the discovery mode for these scalars.

This article is organized as follows: we describe 
our model in
Section~\ref{model}. The most relevant features of the $Hb\bar{b}$
production mechanism in composite Higgs models are discussed in
Section~\ref{HbbinCHM}. We then introduce the experimental analysis to search
for this signature at the LHC. We discuss our results, given in
terms of 
exclusion bounds and discovery limits in
section~\ref{discussion}
and we leave our conclusions for section~\ref{conclusions}. We
describe in Appendix~\ref{appendix:A}  
some technical aspects of the model, including the importance of EWSB
effects in the phenomenology of the lightest fermionic resonances and
give an example of the slow decoupling of heavy partners in models of
strong EWSB. The relevance of $4b$ final states as a discovery channel
for mostly singlet composite scalars is discussed in Appendix~\ref{appendix:B}.

\section{The model\label{model}}

We consider a simplified, two-site~\cite{Bini:2011zb} version of the
minimal composite Higgs
model~\cite{Agashe:2004rs,Agashe:2006at,Contino:2006qr}   
that contains a full description of the bottom
sector. This model captures the mechanism of partial
compositeness and the implications of the global symmetries in the composite
sector. For clarity we neglect non-linear
Higgs couplings due to its pseudo-Nambu-Goldstone nature
(see~\cite{Panico:2011pw,DeCurtis:2011yx,DeSimone:2012fs} 
for a discussion of the corresponding effects). 
In this section we will describe the main relevant features of the
model. Further details can be found in~\cite{Bini:2011zb}.

The model consists of a composite sector, with a global
$SU(3)_c \times SU(2)_L \times SU(2)_R \times U(1)_X$ symmetry plus a
$P_{LR}$ parity that exchanges $SU(2)_L$
and $SU(2)_R$, and an elementary sector,
which contains the SM particles minus the Higgs. Among the composite
resonances, the ones that will play a role in the following are a
color octet vector, $G^c$, transforming in the
$(\mathbf{8},\mathbf{1},\mathbf{1})_0$ representation of the global
symmetry, the 
composite Higgs
\begin{equation}
\mathcal{H}=(\textbf{1},\textbf{2},\textbf{2})_0=\begin{bmatrix}\phi_0^\dagger
& \phi^+ \\ 
- \phi^- & \phi_0 \end{bmatrix},
\end{equation}
and the top and bottom partners
\begin{eqnarray}
\mathcal{Q}&=&(\textbf{3},\textbf{2},\textbf{2})_{2/3}=\begin{bmatrix}
T^c  & T^c_{5/3} \\ B^c & 
T^c_{2/3} \end{bmatrix}, \qquad \tilde{T}^c=(\textbf{3},
\textbf{1},\textbf{1})_{2/3},  \\
\mathcal{Q}^\prime&=&(\textbf{3},\textbf{2},\textbf{2})_{-1/3}
=\begin{bmatrix} B_{-1/3}^c  & T^{\prime\,c} \\ B^c_{-4/3} &
B^{\prime\,c} \end{bmatrix}, \qquad \tilde{B}^c=
(\textbf{3},\textbf{1},\textbf{1})_{-1/3}. 
\end{eqnarray}
The subscript in the name of the quark denotes its electric charge,
given by $Q=T^L_3+Y=T^L_3+T^R_3+X$, with $X$ the charge under the
$U(1)_X$ group. Among the quarks with no subscript, $T^c$, $\tilde{T}^c$ and
$T^{\prime\,c}$ have electric charge $2/3$ and $B^c$, $B^{\prime\,c}$
and $\tilde{B}^c$ have
electric charge $-1/3$. Finally, the superscript $c$ is a reminder
that they belong to the composite sector.

The Lagrangian involving these fields reads
\begin{eqnarray}
\mathcal{L}&=& 
-\frac{1}{2} \mathrm{Tr}[G^e_{\mu\nu} G^{e\,\mu\nu}]
+\frac{1}{2} \left(\frac{g_e M_{G^c}}{g_c}\right)^2 (G^e_\mu)^2
+\bar{q}^e_L \mathrm{i}\cancel{D} q^e_L
+\bar{t}^e_R \mathrm{i}\cancel{D} t^e_R
+\bar{b}^e_R \mathrm{i}\cancel{D} b^e_R \nonumber \\
&& 
-\frac{1}{2} \mathrm{Tr}[G^c_{\mu\nu} G^{c\,\mu\nu}]
+\frac{1}{2} M_{G^c}^2 (G^c_\mu)^2
+\frac{1}{2} \mathrm{Tr}[\partial_\mu \mathcal{H}^\dagger \partial^\mu \mathcal{H}]
- V(\mathcal{H}^\dagger \mathcal{H}) \nonumber \\
&&
+\mathrm{Tr}[\bar{\mathcal{Q}}(\mathrm{i}\cancel{\partial} -
g_c \cancel{G}^c - M_{\mathcal{Q}})\mathcal{Q}]
+\bar{\tilde{T}}^c(\mathrm{i}\cancel{\partial} -
g_c \cancel{G}^c - M_{\tilde{T}^c})\tilde{T}^c\nonumber \\
&&
+\mathrm{Tr}[\bar{\mathcal{Q}^\prime}(\mathrm{i}\cancel{\partial} -
g_c \cancel{G}^c - M_{\mathcal{Q}^\prime})\mathcal{Q}^\prime]
+\bar{\tilde{B}}^c(\mathrm{i}\cancel{\partial} -
g_c \cancel{G}^c - M_{\tilde{B}^c})\tilde{B}^c\nonumber \\
&&
-\Big\{ Y_T \mathrm{Tr}[\bar{\mathcal{Q}}\mathcal{H}]\tilde{T}
+Y_B \mathrm{Tr}[\bar{\mathcal{Q}^\prime}\mathcal{H}]\tilde{B}
\nonumber  \\
&& 
~+ \frac{1}{2}\frac{g_e}{g_c}M_{G^c}^2 G^c_\mu
G^{e\,\mu} 
+\Delta_{L1} \bar{q}_L^e (T^c,B^c)^\mathrm{T}
+\Delta_{L2} \bar{q}_L^e (T^{\prime\,c},B^{\prime\,c})^\mathrm{T}
\nonumber \\
&&
~+\Delta_{tR} \bar{t}_R^e \tilde{T}^c 
+\Delta_{bR} \bar{b}_R^e \tilde{B}^c
+\mathrm{h.c.}\Big\}+\ldots ~.\label{lagrangian}
\end{eqnarray}
The first line involves only elementary fields (denoted with a
superscript $e$), the next four only
composite states and the last two the linear mixing among the two
sectors realizing partial compositeness. This linear mixing can be
eliminated by performing the appropriate rotations so that the
physical particles (before EWSB) are an admixture of
elementary and composite states. For instance we can define the
physical SM gluon and heavy gluon as follows:
\begin{equation}
\begin{pmatrix} G_\mu \\ G_\mu^\ast \end{pmatrix} 
= \begin{pmatrix}
c_s & s_s \\ -s_s & c_s 
\end{pmatrix}
\begin{pmatrix} G^e_\mu \\ G_\mu^c \end{pmatrix}, 
\end{equation}
with $s_s/c_s\equiv \sin \theta_s /\cos \theta_s=g_e/g_c$.
The SM gluon is of course massless and has a coupling $g_s=s_s g_c =
c_s g_e$ and the heavy gluon has a mass $M_{G^\ast}=M_{G^c}/c_s$ and
coupling $-g_s s_s/c_s$ to elementary states and $g_s c_s/s_s$ to
composite ones. In a similar way we can define the SM $t_R$ and $b_R$
and the heavy vector-like singlets $\tilde{T}$ and $\tilde{B}$,
\begin{eqnarray}
\begin{pmatrix} t_R \\ \tilde{T}_R \end{pmatrix} 
&=& \begin{pmatrix}
c_{tR} & -s_{tR} \\ s_{tR} & c_{tR} 
\end{pmatrix}
\begin{pmatrix} t^e_R \\ \tilde{T}_R^c \end{pmatrix}, 
\qquad
\tilde{T}_L = \tilde{T}_L^c,
\\
\begin{pmatrix} b_R \\ \tilde{B}_R \end{pmatrix} 
&=& \begin{pmatrix}
c_{bR} & -s_{bR} \\ s_{bR} & c_{bR} 
\end{pmatrix}
\begin{pmatrix} b^e_R \\ \tilde{B}_R^c \end{pmatrix}, 
\qquad\tilde{B}_L = \tilde{B}_L^c.
\end{eqnarray}
with
\begin{equation}
\frac{s_{tR}}{c_{tR}}=\frac{\Delta_{tR}}{m_{\tilde{T}^c}}, 
\quad 
\frac{s_{bR}}{c_{bR}}=\frac{\Delta_{bR}}{m_{\tilde{B}^c}}, 
\qquad 
M_{\tilde{T}}=\frac{M_{\tilde{T}^c}}{c_{tR}},
\quad 
M_{\tilde{B}}=\frac{M_{\tilde{B}^c}}{c_{bR}}.
\end{equation}
The fact that the SM left-handed doublet mixes with two different
sectors through $\Delta_{L1}$ and $\Delta_{L2}$ complicates the
expressions for the corresponding rotations. They can be found in the
$\Delta_{L2}\ll \Delta_{L1}$ limit in~\cite{Bini:2011zb} and are
reproduced, for further discussion, in Appendix~\ref{appendix:A} for
the charge $-1/3$ sector. This limit
is well motivated by the stringent constraints on the $Zb_L \bar{b}_L$
coupling (which receives corrections that are suppressed by the ratio
$\Delta_{L2}/\Delta_{L1}$), it explains the fact that
$m_b\ll m_t$ and it is naturally generated by the renormalization flow
in the conformal phase~\cite{Contino:2006qr}. In any case we will not
make 
use of the explicit expressions
since in practice we will use the top and bottom quark masses to
(numerically) fix
the values of $\Delta_{L1}$ and $\Delta_{L2}$ in terms of the
remaining parameters of the model. We have checked that, in
all the cases we have considered, the hierarchy
$\Delta_{L2}/\Delta_{L1}\ll 1$ is preserved. 

As we said, the 
top and bottom quark masses are used to fix the values of $\Delta_{L1}$ and
$\Delta_{L2}$. A third parameter, that we take $\theta_s$ can be fixed
from the value of the strong coupling constant
\begin{equation}\sin \theta_s=\frac{g_s}{g_c}.
\end{equation}
All the other
parameters, namely $g_c$, $Y_T$, $Y_B$,
$M_{G^c},M_{\mathcal{Q}},M_{\mathcal{Q}^\prime},M_{\tilde{T}^c},M_{\tilde{B}^c},s_{tR},s_{bR}$,
can be allowed to vary. In order to reduce the dimensionality of the
parameter space we have fixed all the composite fermion
masses~\footnote{Note that these are the masses of the composite
states before EWSB and before their mixing with the elementary
states.} to a
common one
\begin{equation}
M_{\mathcal{Q}}=M_{\mathcal{Q}^\prime}=M_{\tilde{T}^c}=M_{\tilde{B}^c}\equiv
M_F. 
\qquad\mbox{(Universal Masses)}\label{benchmark:1}
\end{equation}
Similarly we have fixed
\begin{equation}
Y_T=Y_B=3,\label{benchmark:2}
\end{equation}
as they are expected to be numbers somewhat larger than one.
\begin{figure}[ht]
{\includegraphics[width=0.48\columnwidth]{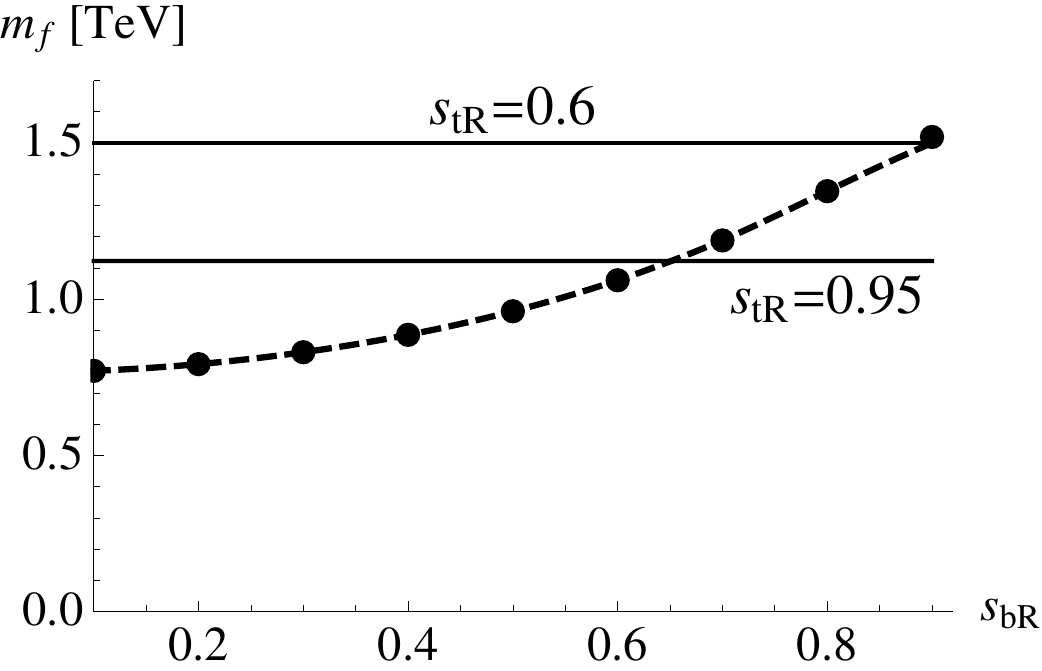} \hfil
\includegraphics[width=0.48\columnwidth]{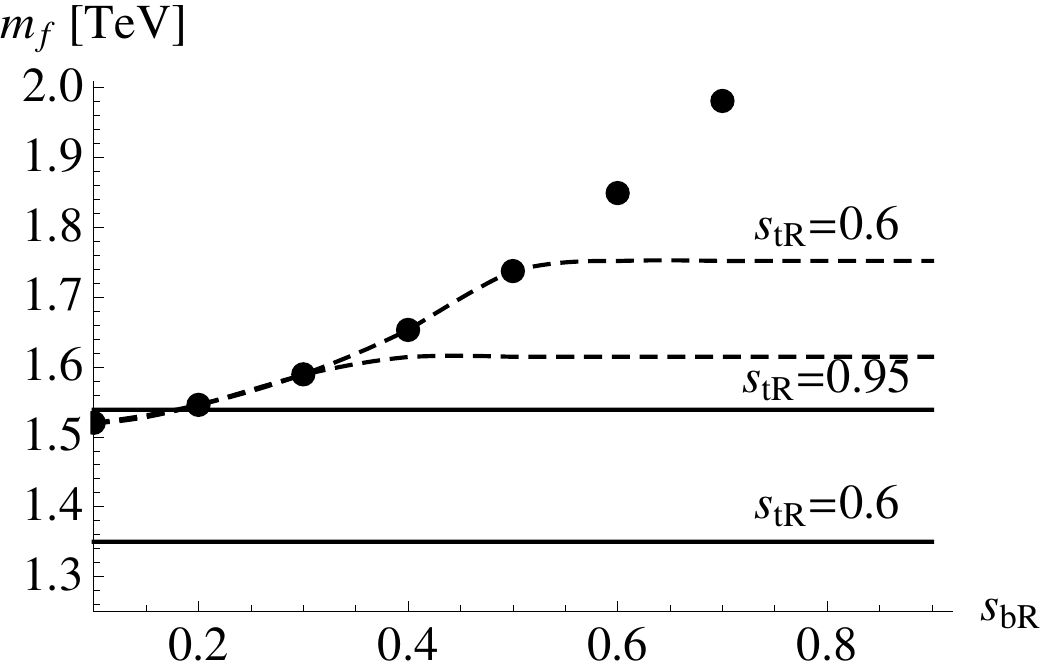}
}
\caption{Mass of the lightest charge $2/3$ (solid) and $-1/3$ (dashed)
quark as a function of $s_{bR}$ and for different values of
$s_{tR}$. The dots correspond to the mass of $B_H$ (see text). In the
left panel we have used Eq. (\ref{benchmark:1}) with $M_F=1.5$ TeV and
in the right one we have used Eq. (\ref{benchmark:1alt}) with
$M_Q=1.5$ TeV.}\label{fig:mlightest}
\end{figure}
For each value of $s_{tR}$, $s_{bR}$ and $M_F$ the fermion spectrum is
then completely fixed. Under the assumption of universal composite
fermion masses, Eq. (\ref{benchmark:1}), the lightest
new fermion is almost always a charge $-1/3$ quark that decays, with
$100\%$ branching ratio, into $Hb$.~\footnote{The $100\%$ branching
fraction into $Hb$ has to do with the degenerate bidoublet structure
of the model, see Appendix~\ref{appendix:A}. If the heavy particle
involved was an electroweak doublet, instead of a custodial bidoublet,
we would get $BR(B_H \to H b)
=BR(B_H \to Z b)=50\%$.} We show in the left panel of
Fig.~\ref{fig:mlightest} the mass of the lightest charge $2/3$ (solid
horizontal lines) and charge $-1/3$ (dashed line) new quarks as a
function of $s_{bR}$ and for two different values of $s_{tR}$,
corresponding to a mildly ($s_{tR}=0.6$) and very strongly
($s_{tR}=0.95$) composite $t_R$,
respectively. We have assumed $M_F=1.5$ TeV (which corresponds to the
mass of the charge $5/3$ and charge $-4/3$ new quarks). The dots in
the figure represent the mass of the charge $-1/3$ new quark that
decays predominantly (with $100\%$ branching ratio for the parameters
in the plot) into $H b$. For a universal fermion mass this always
agrees with the lightest one. Naturalness arguments and the observed
value of the Higgs boson mass typically predict the lightest
fermionic resonances to be the $(T^c_{5/3},T^c_{2/3})$
multiplet~\cite{Matsedonskyi:2012ym,Redi:2012ha,
Marzocca:2012zn,Pomarol:2012qf,Panico:2012uw}. 
In order to test this scenario we have considered
an alternative fermion mass configuration in which all multiplets are
$50\%$ heavier than $Q$,
\begin{equation}
M_{Q^\prime}=M_{\tilde{T}^c}=M_{\tilde{B}^c}=1.5 M_Q. 
\qquad\mbox{(Lightest $Q$)}.\label{benchmark:1alt}
\end{equation}
The resulting spectrum of lightest modes, for $M_Q=1.5$ TeV is shown
in the right panel of Fig.~\ref{fig:mlightest} with the same notation
than in the left panel of the figure. The masses of the charge $5/3$
and $-4/3$ quarks are in this case 1.5 TeV and 2.25 TeV, respectively. 
The mass
of the lightest charge $-1/3$ quark now depends on the degree of
compositeness of $t_R$ and the one decaying predominantly into $Hb$ is
not always the lightest one. Still there is a relatively light charge
$-1/3$ quark with a $100\%$ branching ratio into $Hb$. In the
following we will denote this quark, which is the one we will be
focusing on in this work, $B_H$.
\begin{figure}[t]
{\includegraphics[width=0.55\columnwidth]{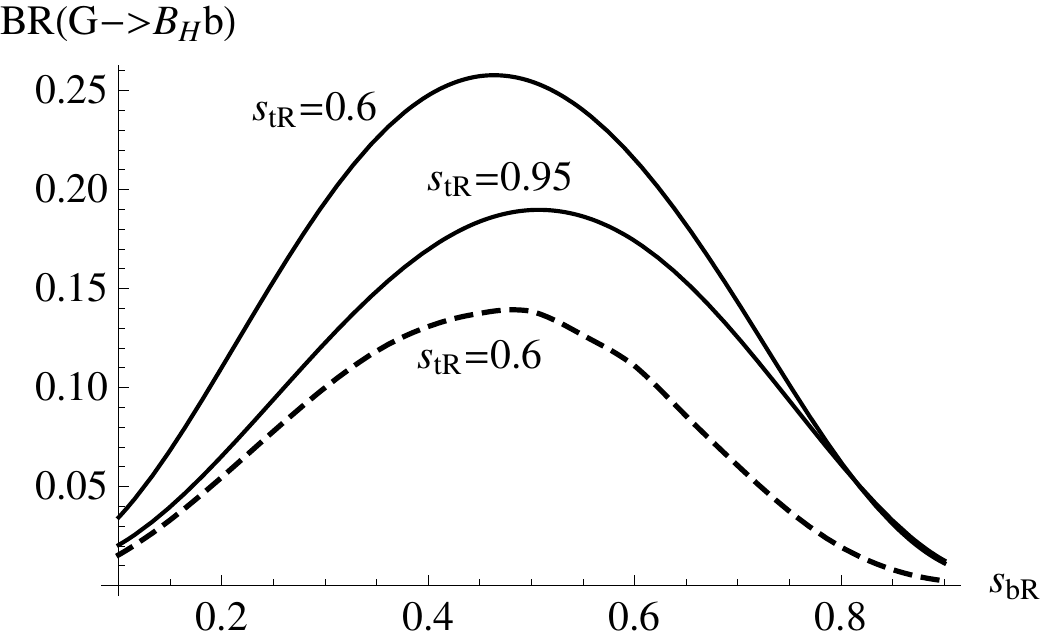}
}
\caption{Branching ratio of the heavy gluon into
$B_H \bar{b}+\bar{B}_Hb$ as a function of $s_{bR}$ and for different
values of $s_{tR}$. We use solid lines for the choice of fermion
masses in Eq. (\ref{benchmark:1}) and dashed lines for
Eq. (\ref{benchmark:1alt}). 
We have fixed $M_{G^\ast}=2.5$ TeV, $g_c=3$ and the
mass of the fermion resonances are fixed so that the lightest new
fermion has a mass $M_{G^\ast}/2$.}
\label{fig:BRGtoBb}
\end{figure}

Once we have discussed the features of the fermionic spectrum and
their decay patterns, we turn our attention to the only two remaining
parameters in the model, namely the heavy gluon, $M_{G^\ast}$, 
mass and the composite
coupling, $g_c$. 
In order to avoid too large a width for the heavy gluon we will choose
its mass so that pair production of top and bottom partners is
kinematically forbidden. Thus, we fix the mass of the heavy
gluon to have twice the mass of the lightest new fermion after EWSB.
In practice what we do is to choose a value for
$M_{G^\ast}$ and fix the value of $M_F$ that makes the mass of the
lightest new fermion $M_{G^\ast}/2$.
Once the value of $M_{G^\ast}$ is fixed, all the phenomenological
implications of the model can be worked out. We show in
Fig.~\ref{fig:BRGtoBb} the branching ratio of the heavy gluon into
$B_H \bar{b}+\bar{B}_H b$ as a function of $s_{bR}$ for different
values of $s_{tR}$. Solid and dashed lines are used for benchmarks
Eq. (\ref{benchmark:1}) and Eq. (\ref{benchmark:1alt}),
respectively. We have fixed $M_{G^\ast}=2.5$ TeV and $g_c=3$ in this
figure. The bell-like shape of the figure arises from the fact that
the coupling between the heavy gluon and $b_R B_{H\,R}$ is
proportional to $s_{bR} c_{bR}$ (see Appendix~\ref{appendix:A}). 
\begin{figure}[t]
{
\includegraphics[width=0.60\columnwidth]{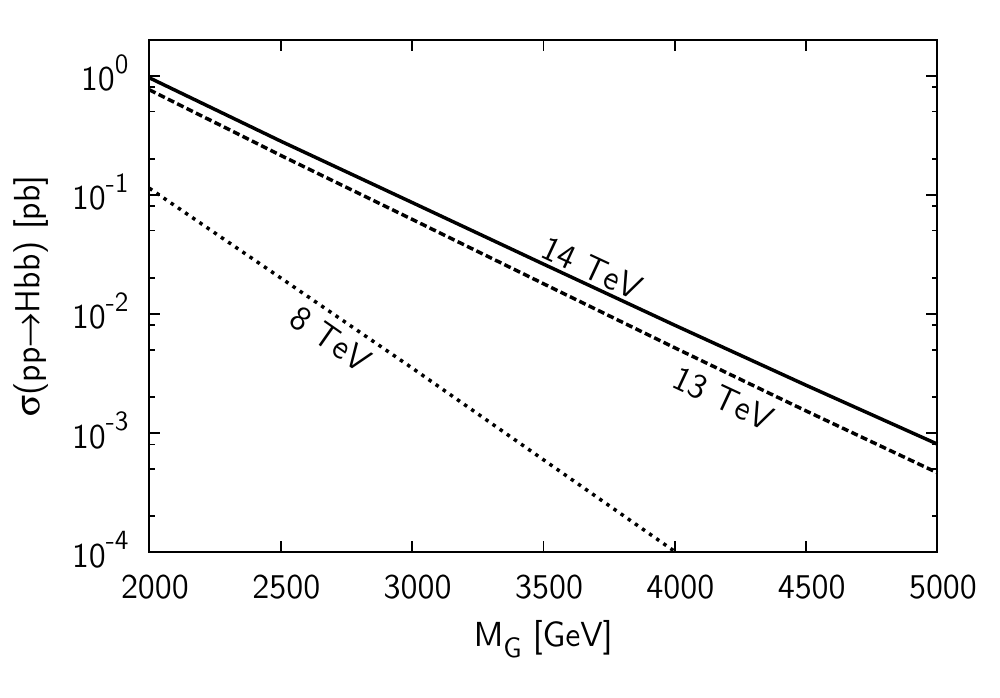}
}
\caption{
$Hb\bar{b}$ production cross section in the
benchmark model, Eqs. (\ref{benchmark:1},\ref{benchmark:2}), with
$g_c=3$, as a 
function of $M_{G^\ast}$. $M_F$ in Eq. (\ref{benchmark:1}) has been
chosen such that the lightest fermionic resonance has a mass
$M_{G^\ast}/2$. 
\label{fig:diagramandxsec}
}\end{figure}
\section{$H b\bar{b}$ via single production of top/bottom
partners \label{HbbinCHM}} 

As we have discussed in the previous section, the heavy gluon can have a
sizeable decay branching ratio into $B_H \bar{b}+\bar{B}_H b$, where
$B_H$ is a charge $-1/3$ quark that is typically relatively light
and decays always to $H b$. Thus,
single production of $B_H$ via the s-channel exchange of $G^\ast$
results in an $H b\bar{b}$ final state with a significant production
cross section. 
We show in Fig.~\ref{fig:diagramandxsec} the $Hb\bar{b}$ 
production cross
section, as a function of the heavy gluon mass, 
with the parameters fixed according to
Eqs. (\ref{benchmark:1}) and (\ref{benchmark:2}), $g_c=3$ 
and $M_F$ chosen such that
the lightest new fermion has a mass equal to $M_{G^\ast}/2$. 
This production cross section is sizeable but not large
enough to allow us to use the cleaner $H\to \gamma \gamma,~ZZ^\ast$  decay
channels. Among the two leading decay channels, we have found that the
$H\to b \bar{b}$ is the most promising one. 
The main reasons are the
large number of $b$ quarks in the final state, which is a very powerful
discriminator against the background, together with very special
kinematics inherited from the large masses of $G^\ast$ and $B_H$. As
we now show, the latter ensures a clean trigger and a very simple
reconstruction algorithm. 

The process we are interested in is therefore
\begin{equation}
pp \to G^\ast \to B_H \bar{b} + \bar{B}_H b \to H b \bar{b} \to 4b.
\end{equation}
Due to the large masses we can probe at the LHC, all
four b quarks in the final state are very hard. 
\begin{figure}[t]
{\hspace{0.5cm}\includegraphics[width=0.45\columnwidth]{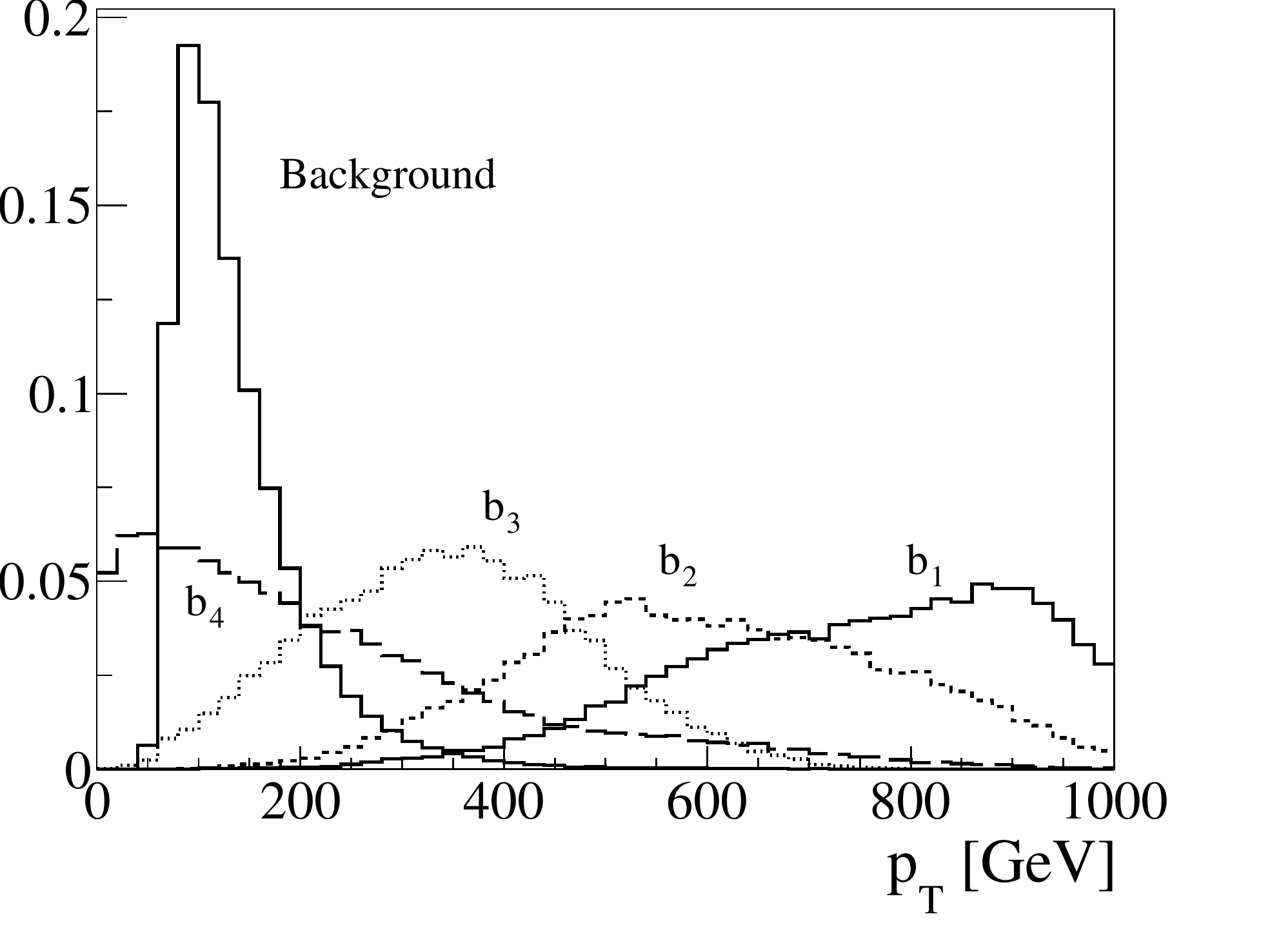}
\raisebox{-.008\height}{\includegraphics[width=0.45\columnwidth]{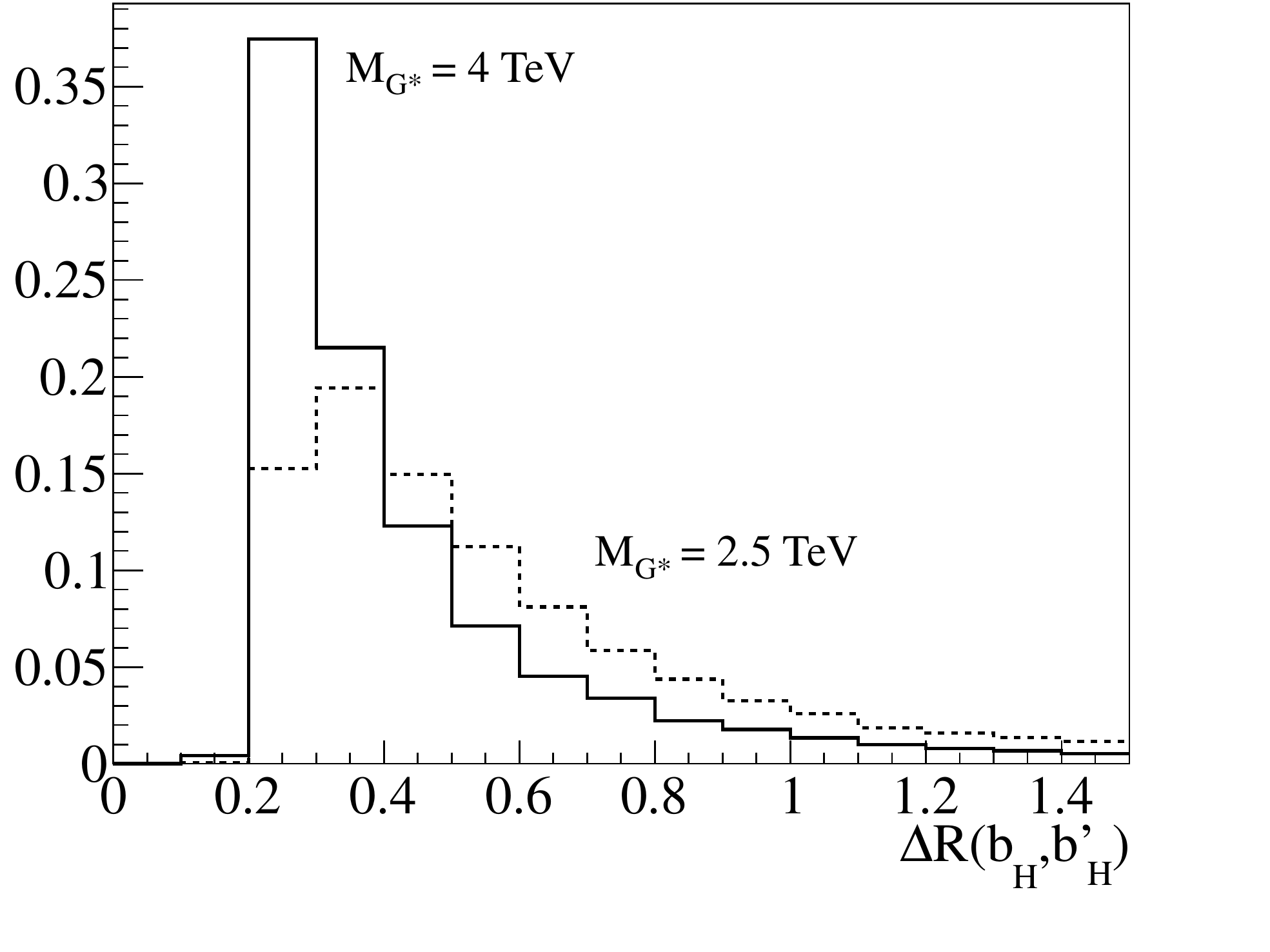}}
}
\caption{Left: Parton level $p_T$ distribution of the 4 $b$ quarks in the
signal (denoted in decreasing order of $p_T$ by $b_{1,2,3,4}$) and of the
hardest $b$ quark in the irreducible $4b$ background. Right: $\Delta R$ separation between the two
$b$-quarks from the Higgs decay at the partonic level for two
different values of the heavy gluon mass. The mass of the heavy bottom is
in both cases $M_{B_H}=M_{G^\ast}/2$. All
distributions are normalized to unit area.}
\label{fig:ptbj} 
\end{figure}
We show in
Fig.~\ref{fig:ptbj} (left) the $p_T$ distribution of the four $b$
quarks at the partonic level, for a heavy gluon mass $M_{G^\ast}=2.5$
TeV, together with the $p_T$ distribution of the hardest $b$ quark for the
irreducible $4b$ QCD background (distributions are normalized to
unit area). All four b-jets are quite
hard with the $p_T$ of the two leading jets well above $300$ and $200$ GeV,
respectively. This allows for a very clean trigger of the signal
events and also for the possibility of hard cuts on the $p_T$ of the
leading $b$-jets, an important ingredient to bring the irreducible
background down to manageable levels.

One important feature is that, due to the
relatively large mass of $B_H$, the Higgs boson tends to be quite
boosted and its decay products relatively aligned. We show in
Fig.~\ref{fig:ptbj} (right) the $\Delta R$ separation between the two $b$-quarks that
reconstruct the Higgs, at the partonic level, for two different values
of $M_{G^\ast}$ (recall that we have $M_{B_H}=M_{G^\ast}/2$). We find
that less than $35\%$ of the events have $\Delta R<0.4$ for
$M_{G^\ast}=2.5$ TeV. This number goes up to $60\%$ for $M_{G^\ast}=4$
TeV. Thus, it is clear that for larger heavy gluon masses, 
the use of boosted
techniques~\cite{Abdesselam:2010pt,Altheimer:2012mn} is likely to enhance the
sensitivity. However, we have decided to restrict ourselves to
traditional techniques 
because the use of one less b-tag would force us to consider
new background processes that are difficult to estimate with other
means than data-driven methods.

\section{Experimental analysis \label{analysis}}

In this section we describe a very simple experimental analysis that
takes advantage of the kinematical features discussed in the
previous section to disentangle the signal from the background. 
In our simulations we have used \texttt{MadGraph
v4} \cite{Alwall:2007st} and \texttt{Alpgen
v2.13} \cite{Mangano:2002ea} for parton level signal and background
generation, respectively. We have set the factorization and
renormalization scales to the default values and used the CTEQ6L1 
PDFs~\cite{Pumplin:2002vw}. 
We have used \texttt{Pythia v6} \cite{Sjostrand:2006za} for
parton showering and hadronization and \texttt{Delphes
v1.9} \cite{Ovyn:2009tx} for fast detector simulation. Jets are
reconstructed using the anti-kt algorithm with $R = 0.4$, and we are
assuming a value of $0.7$ for the b-tagging efficiency.
Jets and charged leptons used in our analysis are defined to have
$p_T^j > 20$ GeV. Charged leptons are also
required to be well isolated from jets with $\Delta R(lj) > 0.4$. 
We have considered two different configurations for the LHC
parameters with benchmark values $\sqrt{s}=8$ TeV and an integrated
luminosity of 20 fb$^{-1}$ (LHC8) and $\sqrt{s}=14$ TeV with an
integrated luminosity of 100 fb$^{-1}$ (LHC14).

The main background to the $p p \rightarrow G^*\rightarrow
B_H\bar{b}+\bar{B}_H b\rightarrow Hb\bar{b}\rightarrow 4b$ process we
are interested in comes from
the irreducible QCD $4b$ production. Other purely hadronic backgrounds
are suppressed by the small b-tagging fake-rate (we conservatively set
$1/100$ for light jets and $1/10$ for c-jets) and can be
neglected. The same happens to other SM processes in which at least
one isolated lepton is produced (we will impose a lepton veto to
reduce these to negligible levels). Thus, the only background we have
to consider is the irreducible one. Still, the QCD $4b$ cross section
is so large that we have been forced to 
generate events in the phase space region defined by $p_T^b > 50$
GeV and $\Delta R(b,b) > 0.3$ to have a large enough sample.
The
cross section in this region of parameter space is 
$\sim 12$ pb. 
In order to ensure enough statistics we have generated a
number of events corresponding to a luminosity of $\sim 1$ ab$^{-1}$. 
In light of the results of NLO
studies \cite{Greiner:2010ci,Bevilacqua:2013taa} 
we have assumed that the shape in the $p_T$ distributions is
well described by our leading order calculations but the total cross
section must be corrected with a k-factor that we conservatively set
to 1.5. 

In order to bring the irreducible background down to manageable
levels, we impose the following set of cuts:
\begin{eqnarray}
&& N_b\geq 4, \quad N_l=0, \quad 
p_T(b) \geq \left\{ 
\begin{array}{l}
50\mbox{ GeV (LHC8)}, \\
60\mbox{ GeV (LHC14)}, 
\end{array}\right .
\nonumber \\
&& 
p_T(b_1) \geq \left\{ 
\begin{array}{l}
200\mbox{ GeV (LHC8)}, \\
300\mbox{ GeV (LHC14)}, 
\end{array}\right .
\quad
p_T(b_2) \geq \left\{ 
\begin{array}{l}
100\mbox{ GeV (LHC8)}, \\
200\mbox{ GeV (LHC14)}, 
\end{array}\right .
\nonumber \\
&&|m_{b_H b_H^\prime}-m_H|\leq 30\mbox{ GeV}, \label{firsthcuts}
\end{eqnarray}
where we have denoted $b_{1,2,\ldots}$ the $b$-jets in decreasing
order in $p_T$, $b$ generically denotes all $b$-jets and finally $b_H$
and $b_H^\prime$ are the two $b$-jets that better reconstruct the Higgs.
We impose different cuts on the $p_T$ of the $b$-jets for LHC8 and
LHC14.
We now use the
invariant mass of the four leading $b$-jets as the discriminating
variable. We require the events to have a $4b$ invariant mass close to
the test mass for the heavy gluon:
\begin{equation}
M_{G^*}+1000\,\text{GeV}<m_{4b}<M_{G^*} + 500 \,\text{GeV}.\label{secondcut}
\end{equation}
\begin{table}[t]
\begin{center}
\begin{tabular}{p{0.14\columnwidth}|p{0.11\columnwidth}|p{0.11\columnwidth}|p{0.11\columnwidth}|p{0.11\columnwidth}|p{0.11\columnwidth}|p{0.12\columnwidth}|p{0.10\columnwidth}}
\textbf{8 TeV} & $N_b  $ & $N_l  $ & $ p_T^b$  & $ p_T^{b_1}$ &
$p_T^{b_2}$ & $|m_{bb}-m_H|$ & $m(4b)$\\\hline
Signal & 16 & 99 & 68 & 99 & 99 & 56 & 89\\
Background & 17 & 99 & 10 & 13 & 89 & 46 & 0.7\\\hline
\textbf{14 TeV} &  &    &   &  &  &  & \\\hline
Signal & 16 & 99 & 59 & 98 & 98 & 59 & 92\\
Background & 20 & 99 & 12 & 7.6 & 63 & 36 & 11\\\hline
\end{tabular}
\end{center}
\caption{Cut by cut efficiencies (in percent) for the signal in the benchmark model with $M_{G^*} = 2.5$ TeV for two different center of mass energies, and the irreducible $b\bar{b}b\bar{b}$ background. The slightly low efficiency in $N_b$ for the signal is consequence of the boosted regime.}
\label{tab:hefficiencies}
\end{table}
\begin{figure}[t]
\includegraphics[width=0.49\columnwidth]{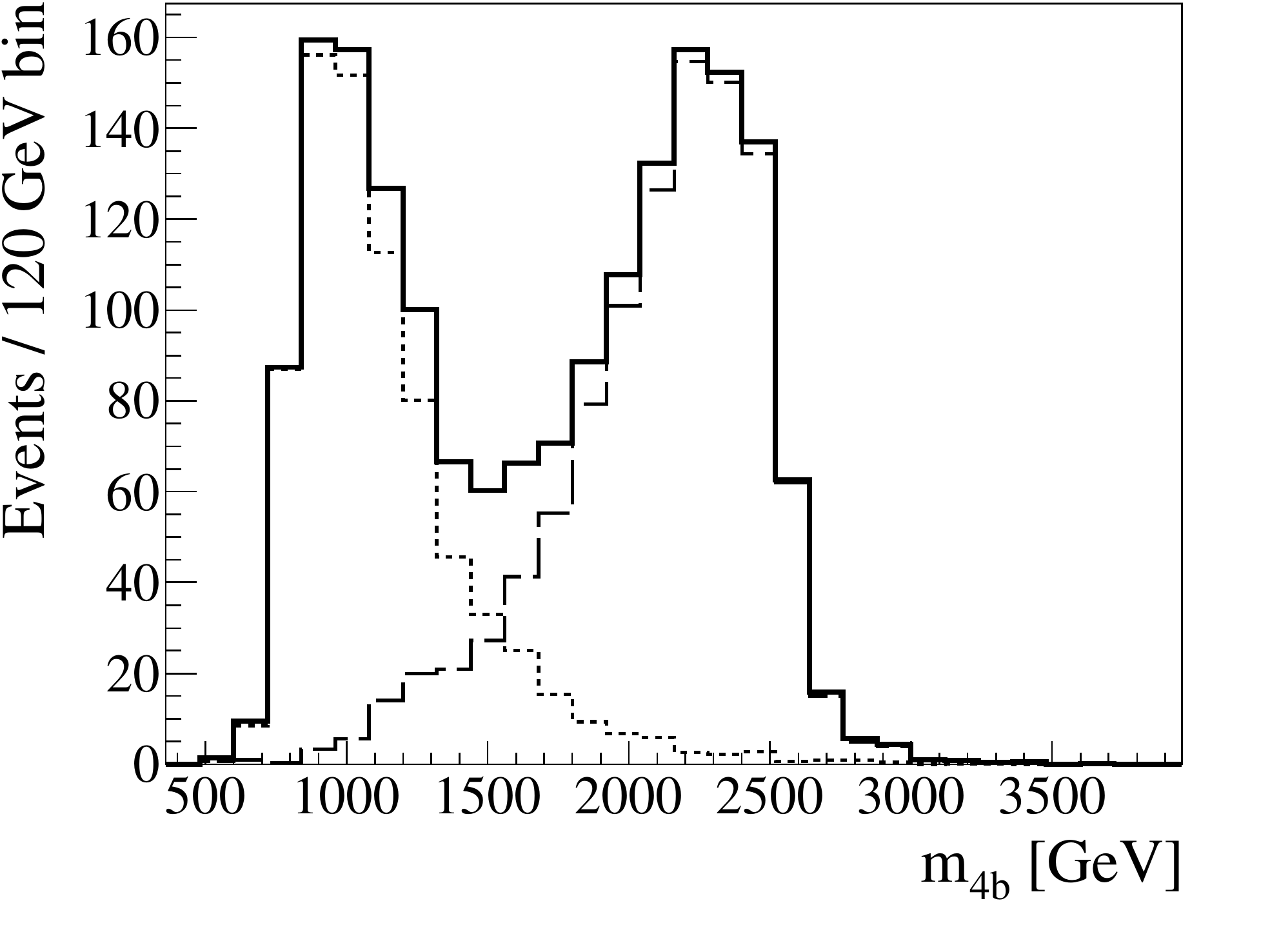}
\includegraphics[width=0.49\columnwidth]{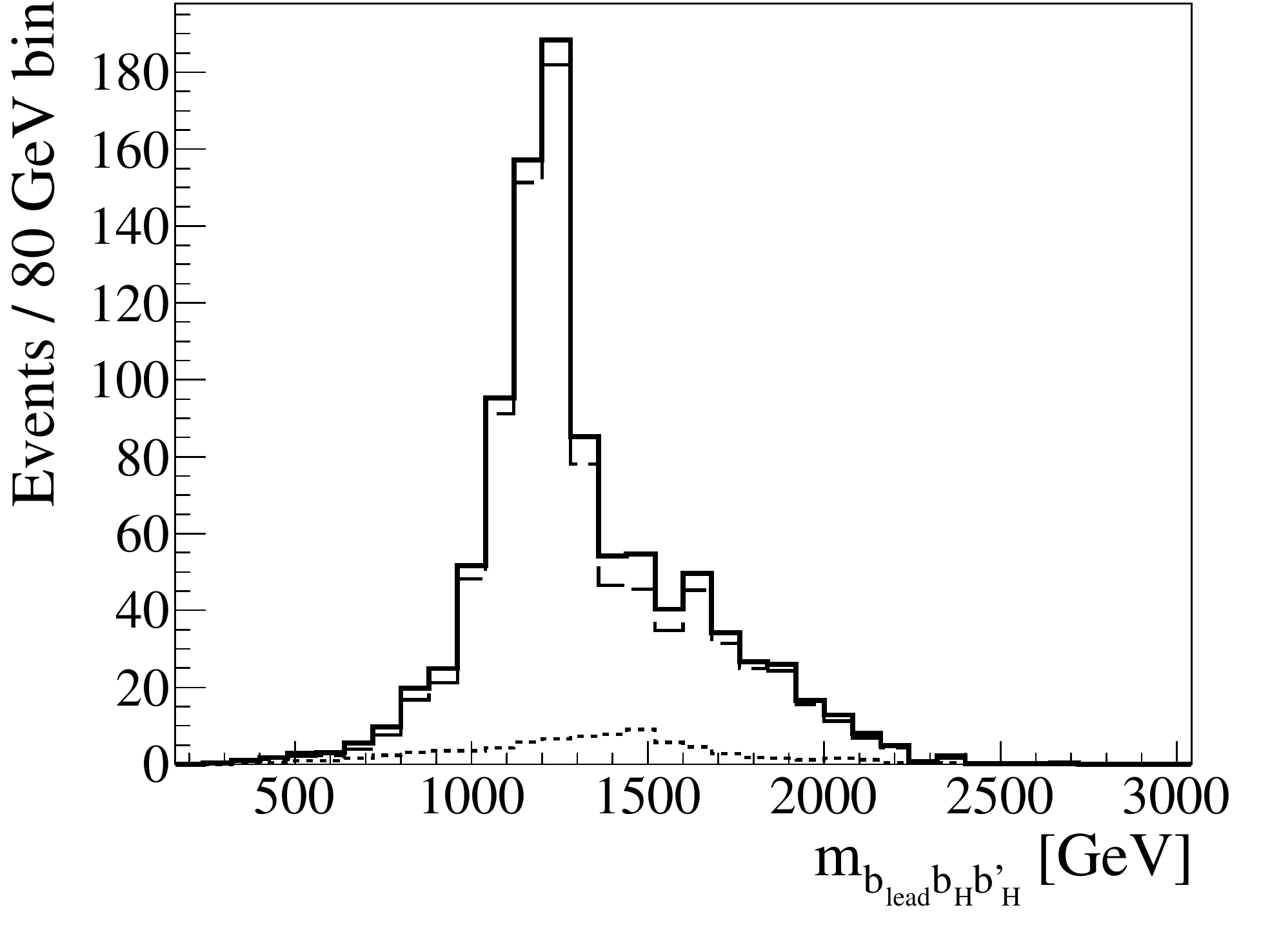}
\caption{Plots of reconstructed events after the cuts of equation
(\ref{firsthcuts}). The dashed, dotted and solid lines represent the
signal, the background and the sum (data) respectively. Left)
reconstruction of $G^*$ from the four leading b-tagged jets. 
Right) reconstruction of
$B_H$ from the two jets reconstructing the Higgs plus the hardest
among the remaining $b$-jets after the cuts in Eqs. (\ref{firsthcuts})
and (\ref{secondcut}). 
}
\label{figmasses}
\end{figure}
The efficiencies of the different cuts for the signal (with
$M_{G^\ast}=2.5$ TeV) and
the irreducible background are given in
Table~\ref{tab:hefficiencies}. 
The relatively low
efficiency for the signal of the $N_b$ cut is due to the fraction of boosted
events.
We show in Fig.~\ref{figmasses} (left) the invariant mass of the four
leading $b$-jets, after the cuts
in Eq. (\ref{firsthcuts}) have been imposed, for the signal and
irreducible background. The figure shows that 
this observable is clearly a discriminating
variable, with a distinct peak around the mass of the heavy
gluon. Cutting on a window around the test mass, the background
is reduced to negligible levels.
Once we have reconstructed the heavy gluon mass, we can reconstruct
the heavy bottom by taking the invariant mass of the two $b$-tagged
jets that best
reconstruct the Higgs mass ($b_H$ and $b_H^\prime$) 
and the leading one among the remaining
$b$-jets (denoted $b_\mathrm{lead.}$). 
We have checked that the peak in this distribution around
the heavy bottom mass is narrower than the one obtained with other
combinations of $b$-jets, for the values of $M_{B_H}$ and $M_{G^\ast}$ we have considered. An example of this is shown in the right
panel of
Fig.~\ref{figmasses}.

\section{Discussion of the results\label{discussion}}

The analysis described in the previous section takes full
advantage of the kinematical features of the signal to extract it from
the background. Other searches, not specifically aimed at this model
can be somewhat sensitive to the signal we are considering. Among
them, the two most important ones are searches with many $b$-quarks in
the final state, typically motivated by supersymmetric models, and
searches for new physics in dijet final states. The latter has been
shown to impose stringent constraints on these kind of
models~\cite{Carmona:2012my} but they are less related to the
particular final state that we are considering in this work. We have
found that, among the former, searches for $Hb\bar{b}$ production in
supersymmetric models and searches for multi-$b$ final states in
association with missing energy are the most sensitive ones. Let us
discuss them in turn.

Searches for $Hb\bar{b}$ (or $Hb$) in supersymmetry look for events
with three or more relatively hard $b$-jets in the final state and try
to reconstruct the Higgs from the two leading $b$-jets. The expected
$p_T$ distribution of the signal in
supersymmetric models is much softer than in our model and therefore
the focus is in a highly background populated region in which our
signal gets easily diluted. This fact, combined with the small luminosity, makes these searches not very
sensitive to our model, although a very simple extension of the
analysis with harder cuts on the $p_T$ of the $b$-tagged jets would
make them a very sensitive probe of composite Higgs models.

Searches for multi-$b$ final states in association with missing
energy, on the other hand, look for signatures with many $b$-jets in
the final state, with a large value of $H_T$ (scalar sum of all the
$b$-jet $p_T$) and a sizeable amount of missing transverse energy. Due
to the large energy of the final state particles in our model, the
fake missing transverse energy is non-negligible and these searches
are sensitive to our model. It is interesting to note that analyses
in which sophisticated observables are used to avoid contamination
from fake missing $E_T$ (like $\alpha_T$ in~\cite{Chatrchyan:2013lya}) 
kill our signal
together with the multi-jet background. 
However, other analyses in which the
rejection of fake missing energy is less sophisticated impose some
constraints on the parameter space of our model. We have
used~\cite{CMS-PAS-SUS-12-024} that analyzes the full 8 TeV LHC data
and show that, although this search imposes some constraints on the
model,  
our modified analysis in which the
missing energy requirement is replaced for a more stringent
requirement in terms of the $p_T$ of the different $b$-jets, leads to
a much better reach. 
This is an example of a very simple modification of current analyses that
could maximize the number of models the searches are sensitive to.

\begin{figure}[ht]
\includegraphics[width=0.49\columnwidth]{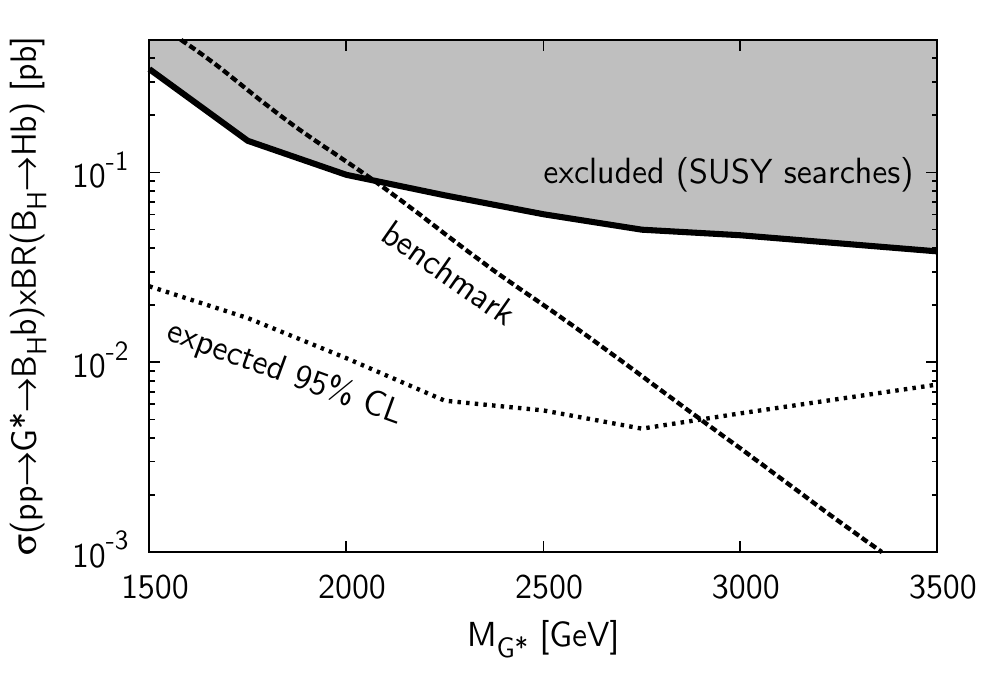}
\hfil
\includegraphics[width=0.49\columnwidth]{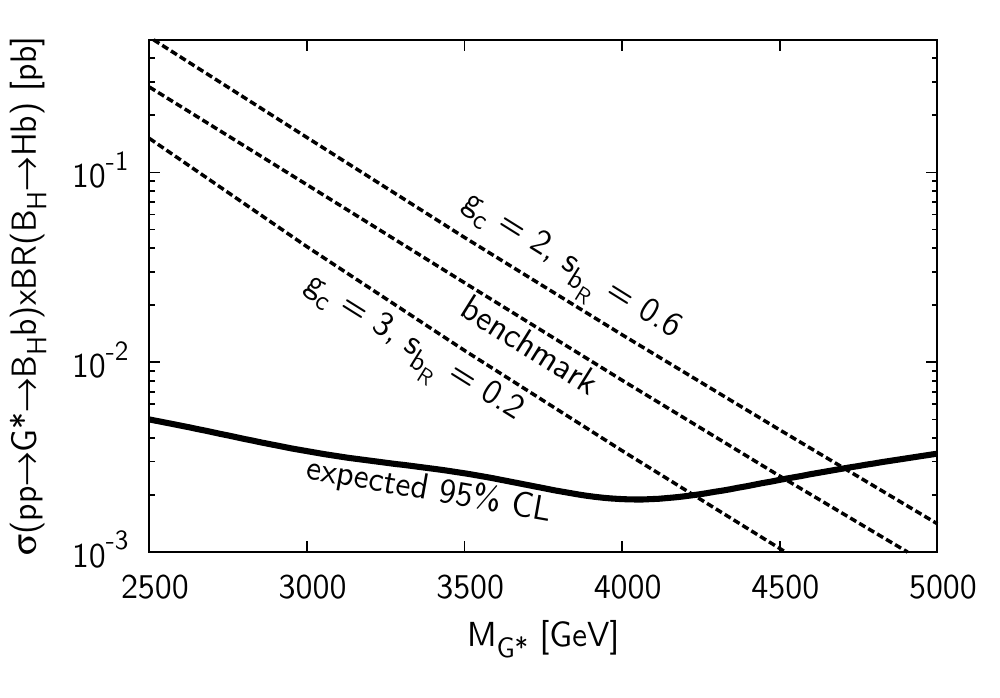}
\caption{$95\%$ C.L. exclusion bound on the $Hb\bar{b}$ production
cross section as a function of the heavy gluon mass for the LHC8 
(left) and LHC14 (right) with 20 fb$^{-1}$ and 100 fb$^{-1}$ 
of integrated luminosity, respectively. The dashed lines correspond to
the cross section in our model for different values of the input
parameters.}
\label{fig:sigmaxbr}
\end{figure}
Once we have described the experimental analysis and our results for
the corresponding efficiencies we can report on the expected bounds
and discovery reach at the LHC. Our main result, summarized in
Fig.~\ref{fig:sigmaxbr}, shows the expected $95\%$ C.L. upper limit on the
$Hb\bar{b}$ production cross section as a function of the heavy gluon
mass. We overlay the cross sections for several points in parameter
space for our model that allow us to compute the corresponding bounds
on $M_{G^\ast}$. The results for the LHC8 are shown in the left panel
of the figure in which we also show the corresponding bound from
current searches on multi-$b$ plus missing energy final states. As we
see, our modified analysis can improve the current limits (using the
same data) 
by more than an order of magnitude in cross section and by almost 1 TeV
in the reach of the heavy gluon mass up to $\sim 3$ TeV for the
benchmark model. 
The expected bound for the
LHC14, together with several different models is shown in the right
panel of the figure. In this case 100 fb$^{-1}$ of integrated
luminosity would allow us to probe masses in the $4-5$ TeV region for
the heavy gluon, depending on the model parameters.

\begin{figure}[ht]
\includegraphics[width=0.49\columnwidth]{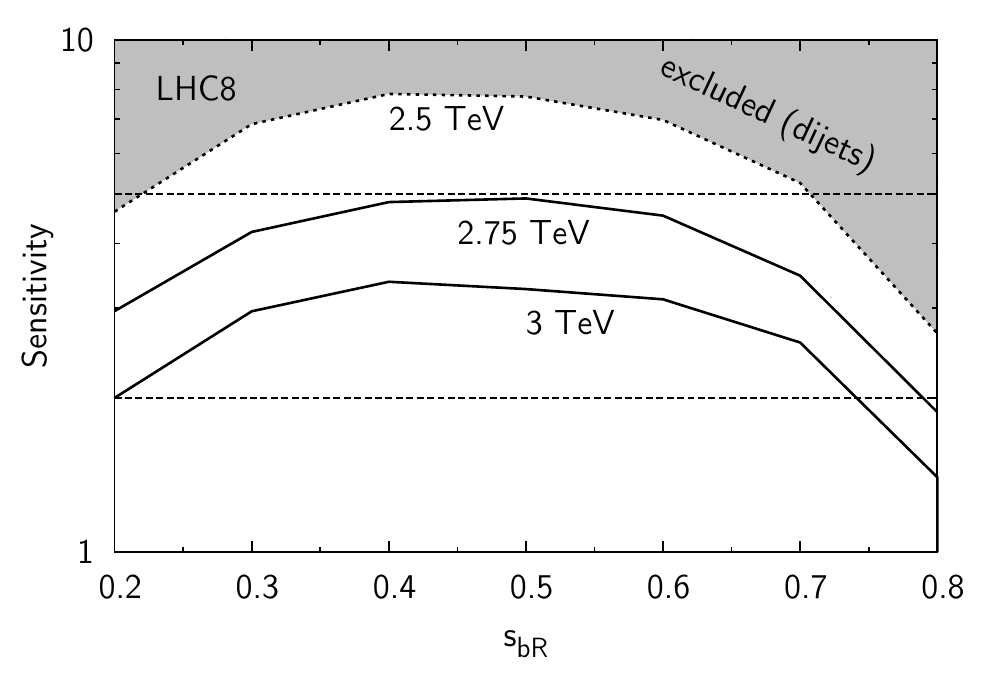}
\hfil
\includegraphics[width=0.49\columnwidth]{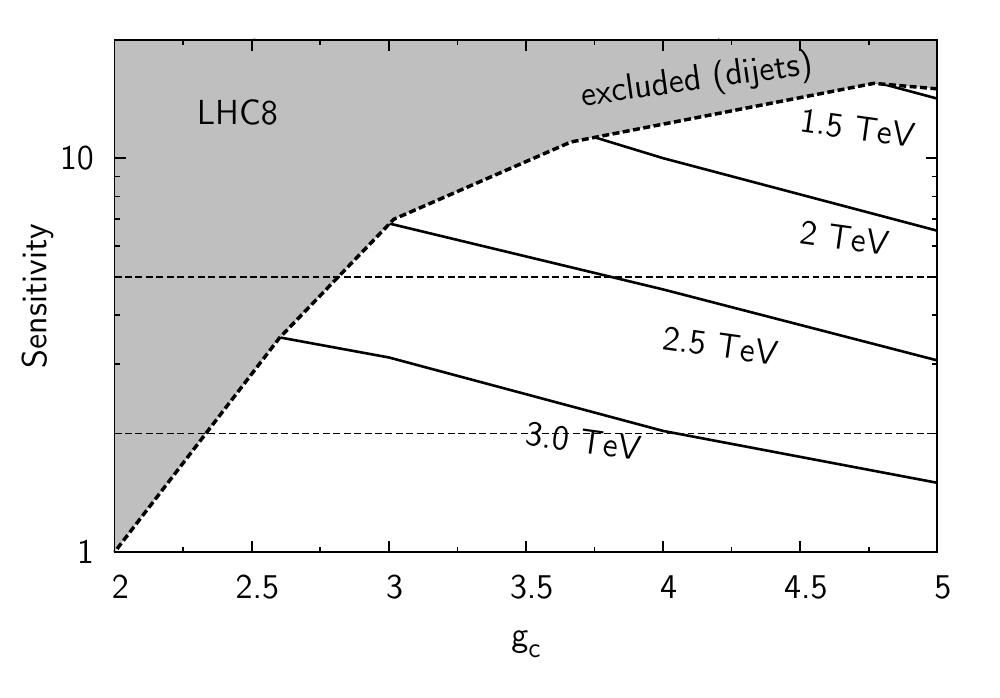}
\\
\includegraphics[width=0.49\columnwidth]{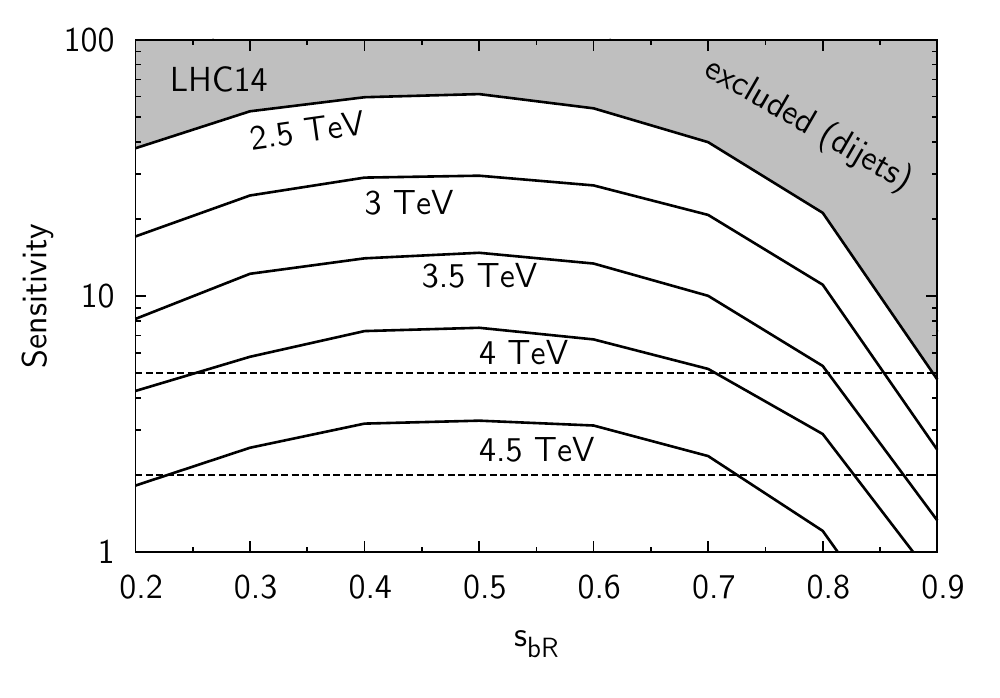}
\hfil
\includegraphics[width=0.49\columnwidth]{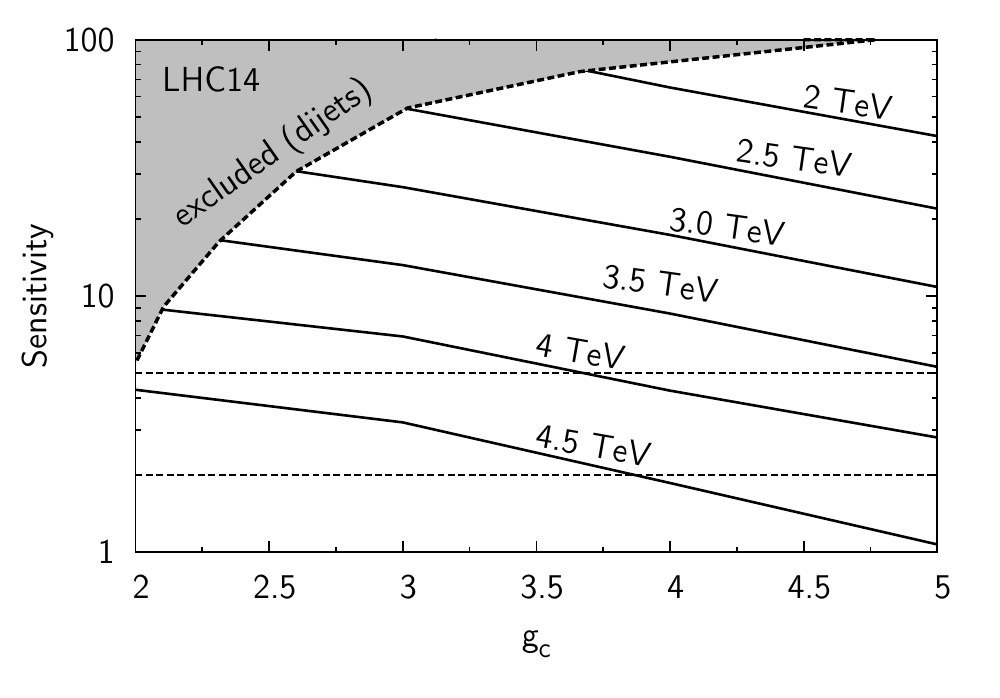}
\caption{Sensitivity reach in the model as a function of $s_{bR}$
(left) and $g_c$ (right), for the LHC8 with 20 fb$^{-1}$ (top) and the
LHC14 with 100 fb$^{-1}$ (bottom). The bounds from current dijet searches are
also shown. 
}
\label{fig:sensitivity}
\end{figure}

The sensitivity of the
LHC8 and LHC14 to different parameters in the model is shown in
Fig.~\ref{fig:sensitivity} in which we give the sensitivity that can
be reached, as a function of $s_{bR}$ (left) and $g_c$ (right), for
different values of the heavy gluon mass and for the two LHC
configurations with LHC8 on the top row and LHC14 on the bottom one. In order to account for the finite statistics, we use \texttt{SigCalc}~\cite{sigcalc}, that takes $\tau \equiv \mathcal{L_\text{MC}}/\mathcal{L}_{\text{data}}$ as an input, where $\mathcal{L_\text{data}}$ and $\mathcal{L}_\text{MC}$ represent the actual and the generated luminosities respectively. The results given by \texttt{SigCalc} reduces to
\begin{equation}
\mathcal{S}(s,b)=\sqrt{2\left((s+b)\log{\left(1+\frac{s}{b}\right)-s}\right)}
\end{equation}
in the limit $\tau\rightarrow\infty$.
In this plot we also show
the bounds derived from dijet searches~\cite{Domenech:2012ai}, which
are more constraining than multi-$b$ searches for our model.
As we see,
despite the stringent bounds on the model from dijet searches, there
are allowed regions in parameter space with heavy gluon masses in the
$1.5-2.75$ TeV range that could be discovered with the LHC8 data.
At the LHC14 masses up to $5$ TeV can be constrained and up to
$4.5$ TeV discovered with 100 fb$^{-1}$.

\section{Conclusions\label{conclusions}}

Light top and bottom partners are a common prediction of natural
models of strong electroweak symmetry breaking. They are new
vector-like quarks that play a direct role in the way the top and
bottom quarks acquire their mass (and their partners under the global 
symmetries of the composite sector). Among these top and bottom
partners, we can have a charge $-1/3$ new quark, heavy bottom, that
decays predominantly into $H b$. The single production of this
heavy bottom via the s-channel exchange of a heavy gluon, a color octet
vector resonance that is also a common prediction in these
models, results in a sizeable $H b\bar{b}$ production. Contrary to
what happens in supersymmetric models, this large 
$Hb\bar{b}$ production cross section is not related to an
enhancement of the bottom quark Yukawa coupling but to the strong
couplings among resonances of the composite sector. Also, in our case,
the relatively large masses of the intermediate states ensures that we
can use the leading $H\to b\bar{b}$ Higgs decay, as all the particles
in the final state are quite hard. This guarantees a clean trigger and
a powerful discriminating power against the large QCD background. 

We have shown that simple modifications of current multi-$b$ final
state searches, typically
motivated by supersymmetric models, can turn theses analyses into very
powerful probes of composite Higgs models. The requirement of very
stringent cuts on the $p_T$ of the different $b$-tagged jets and a
relaxation in the amount of missing energy requested can significantly
reduce the background without sensibly affecting our signal. In this
way, masses up to $\sim 3$ ($2.75$) TeV for the heavy gluon can be excluded
(discovered) with current data at the LHC8. The bounds and discovery
limits go up to $\sim 5$ and $4.5$ TeV, respectively at the LHC14 with 100
fb$^{-1}$ of integrated luminosity. 

We have also shown that electroweak symmetry breaking effects can
substantially modify the collider phenomenology of bottom
partners. The small coupling between the latter and the bottom quark,
together with the very large coupling among composite states can lead
to a puzzling situation from the experimental point of view in which
heavy states, well above the LHC reach, have a profound impact on the
phenomenology of new discovered particles. This effect is explained in
detail in Appendix~\ref{appendix:A} and shows that simplified
models can be a good first approach to new physics searches but they can
also miss some of the main phenomenological properties of realistic
models of new physics. This is particularly true on models of strong
EWSB, in which large couplings among heavy states are naturally
expected.

\begin{acknowledgments}
We would like to thank F. del \'Aguila, G. Azuelos and 
N. F. Castro for useful comments. This
work has been supported by MICINN projects FPA2006-05294 and
FPA2010-17915, through the FPU programme and by Junta de
Andaluc\'{\i}a projects FQM 101, FQM 03048 and FQM 6552.
\end{acknowledgments}

\appendix

\section{Slow decoupling of heavy partners in models of strong
electroweak symmetry breaking \label{appendix:A}}

In this appendix we will explain why the lightest charge $-1/3$ new
quark decays predominantly in the $Hb$ channel. We will also show that
the features of models of strong EWSB with partial compositeness can
lead to the following situation: a single new quark, with electric
charge $-1/3$, is found at the LHC but its decay pattern differs
substantially from the one expected for a vector-like singlet quark.
This is most
easily understood in the basis in which we have diagonalized the mass matrix
before EWSB. As we mentioned in the text, the presence of
$\Delta_{L1}$ and $\Delta_{L2}$ makes this diagonalization
non-trivial. Approximate analytic expressions can be obtained in
the limit $\Delta_{L2}\ll M$, with $M$ any of the dimensionful
parameters in the Lagrangian of Eq. (\ref{lagrangian}). 
This limit is well
motivated by the fact that corrections to the $Z b_{L} \bar{b}_L$
coupling scale like $\Delta_{L2}^2$ and experimental bounds on this
coupling therefore imply that $\Delta_{L2}\ll M$. Furthermore, the
bottom quark mass is also proportional to $\Delta_{L2}$ and we can
relate the absence of large corrections to the $Z b_L \bar{b}_L$
coupling with the fact that $m_b \ll m_t$. Finally, the choice
$\Delta_{L2}\ll \Delta_{L1}$ is radiatively stable. 

The mass matrix for charge $-1/3$ quarks in the 
$b,\tilde{B},B^\prime,B_{-1/3}, B$ basis reads~\cite{Vignaroli:2011um}
\begin{equation}
\mathcal{M}_{-\frac{1}{3}}=\begin{pmatrix}
\frac{v}{\sqrt{2}} Y_B s_2 s_{bR} & - \frac{v}{\sqrt{2}} Y_B s_2
c_{bR} & 0 & 0 & 0 
\\
0 & \frac{M_{\tilde{B}^c}}{c_{bR}} & \frac{v}{\sqrt{2}}Y_B 
 & \frac{v}{\sqrt{2}}Y_B & \frac{v}{\sqrt{2}}Y_B s_4
\\
-\frac{v}{\sqrt{2}} Y_B s_{bR} & \frac{v}{\sqrt{2}} Y_B c_{bR}
& M_{\mathcal{Q}^\prime} & 0 & 0 \\
-\frac{v}{\sqrt{2}} Y_B s_{bR} & \frac{v}{\sqrt{2}} Y_B c_{bR}
& 0 & M_{\mathcal{Q}^\prime}  & 0 \\
-\frac{v}{\sqrt{2}} Y_B s_{bR} s_3 & \frac{v}{\sqrt{2}} Y_B c_{bR} s_3
&0 & 0 & \sqrt{M_{\mathcal{Q}}^2+\Delta_{L1}^2} 
\end{pmatrix}+\mathcal{O}(\Delta_{L2}^2/M^2),\label{M1}
\end{equation}
where
\begin{eqnarray}
s_2 & \equiv & \Delta_{L2}\frac{
M_{\mathcal{Q}}}{M_{\mathcal{Q}^\prime}\sqrt{\Delta_{L1}^2 
+ M_{\mathcal{Q}}^2}} ,
\\
s_3 & \equiv & \Delta_{L2}\frac{ \Delta_{L1}
M_{\mathcal{Q}^\prime}}{(\Delta_{L1}^2+M_{\mathcal{Q}}^2 -
M_{\mathcal{Q}^\prime}^2)\sqrt{\Delta_{L1}^2
+ M_{\mathcal{Q}}^2}}, 
\\
s_4 & \equiv & \Delta_{L2}\frac{\Delta_{L1}}{
\Delta_{L1}^2+M_{\mathcal{Q}}^2 - M_{\mathcal{Q}^\prime}^2}.
\end{eqnarray}
Note that $s_{2,3,4}$ are all proportional to $\Delta_{L2}$ and are
therefore expected to be small.
The fields of this basis are written in terms of the elementary and
composite states as follows:
\begin{align}
&b_L = c_1 b_L^e-s_1 B_L^c - s_2 B_L^{\prime\,c}, 
&b_R = c_{bR} b_R^e - s_{bR} \tilde{B}_R^c, \\
&B_L = s_1 b_L^e + c_1 B_L^c+s_3 B_L^{\prime\,c}, 
&B_R = B_R^c + s_4 B_R^{\prime\, c}, \\
&B_L^\prime = (s_2 c_1 -s_1 s_3)b_L^e - (c_1 s_3+s_1s_2)B_L^c +
B_L^{\prime\,c},
&B_R^\prime = B_R^{\prime\,c} - s_4 B_R^c, \\
&B_{-1/3\,L} =B_{-1/3\,L}^c, &B_{-1/3\,R} = B_{-1/3\,R}^c\\
&\tilde{B}_L = \tilde{B}_L^c, 
&\tilde{B}_R = s_{bR} b_R^e + c_{bR} \tilde{B}_R^c, 
\end{align}
where
\begin{equation}
s_1=\frac{\Delta_{L1}}{\sqrt{\Delta_{L1}^2+M_{\mathcal{Q}}^2}},
\end{equation}
and $c_i=\sqrt{1-s_i^2}$ for $i=1,\ldots,4$.
In this basis, the heavy gluon has the following off-diagonal
couplings
\begin{equation}
\mathcal{L} = \frac{g_s}{\sin \theta_s \cos \theta_s} 
G^\ast_\mu 
\Big[s_1 c_1 \bar{b}_L \gamma^\mu B_L  + s_{bR}
c_{bR} \bar{b}_R \gamma^\mu \tilde{B}_R + \mathrm{h.c.}\Big]+\ldots,
\end{equation}
where we have neglected terms that are suppressed by $\Delta_{L2}/M$.
The mass matrix in Eq. (\ref{M1}) can be further simplified by means
of the following rotation
\begin{equation}
B_{L,R}^\pm = \frac{\pm B^\prime_{L,R} + B_{-1/3\,L,R}}{\sqrt{2}}.
\end{equation}
In the new basis $b,\tilde{B},B^+,B^-,B$, the mass matrix reads
\begin{equation}
\mathcal{M}_{-\frac{1}{3}}^\prime =\begin{pmatrix}
\frac{v}{\sqrt{2}} Y_B s_2 s_{bR} & - \frac{v}{\sqrt{2}} Y_B s_2
c_{bR} & 0 & 0 & 0 
\\
0 & \frac{M_{\tilde{B}^c}}{c_{bR}} & v Y_B 
 & 0 & \frac{v}{\sqrt{2}}Y_B s_4
\\
-v Y_B s_{bR} & v Y_B c_{bR}
& M_{\mathcal{Q}^\prime} & 0 & 0 \\
0 & 0
& 0 & M_{\mathcal{Q}^\prime}  & 0 \\
-\frac{v}{\sqrt{2}} Y_B s_{bR} s_3 & \frac{v}{\sqrt{2}} Y_B c_{bR} s_3
&0 & 0 & \sqrt{M_{\mathcal{Q}}^2+\Delta_{L1}^2} 
\end{pmatrix}.\label{M2}
\end{equation}
Recall that $s_{2,3,4}$ are expected to be small. 
In this case, all the Yukawa couplings inducing
mixing among the different quarks are suppressed except for the ones
between $\tilde{B}$ and $B^+$, which are large (recall that we expect
$Y_B \sim \mathcal{O}(\mbox{ a few})$), and the coupling of $B^+_L$
and $b_R$ with will again be unsupressed except for very small values
of $s_{bR}$. Under the assumption of universal masses,
$M_{\tilde{B}^c}=M_{\mathcal{Q}}=M_{\mathcal{Q}^\prime}$, the lightest
charge $-1/3$ quark is then a combination of $\tilde{B}$ and $B^+$
that inherits the sizeable coupling to the heavy gluon from its
$\tilde{B}$ component and the overwhelming decay into $H b$ from its
$B^+$ component.~\footnote{$B^\pm$ are the symmetric and antisymmetric
combinations of quarks with third component of isospin $T^3_L=\pm 1/2$,
respectively. In the absence of any further mixing they have the
following decay pattern $BR(B^+\to H b)=BR(B^-\to
Zb)=1$. See~\cite{Atre:2008iu,delAguila:2010es,Atre:2011ae,Atre:2013ap} 
for a detailed discussion and collider implications.}

The large coupling between $B^+$ and $\tilde{B}$, together with the
suppressed coupling between $\tilde{B}$ and the SM bottom quark can
lead to an interesting situation in which the heavy partners show a
very slow decoupling. In order to see this effect, let us consider the
limit in which the only light new quark is $\tilde{B}$
\begin{equation}
v \ll M_{\tilde{B}^c} \ll M_{\mathcal{Q}} \sim
M_{\mathcal{Q}^\prime},\quad \mbox{ (slow decoupling)}.\label{slow:decoupling}
\end{equation} 
In particular we consider that $\tilde{B}$ is well within the LHC
reach whereas all other particles are well above the LHC threshold.
The small value of $s_3$ and $s_4$ allow us to disregard the $B^-$ and
$B$ fields in the following and focus the discussion in
$b$,$\tilde{B}$ and $B^+$. In the limit in which
$M_{\mathcal{Q}^\prime}\to \infty$, $\tilde{B}$ is an electroweak
singlet and will have the standard decay pattern
\begin{equation}
BR(\tilde{B}\to W t)/ BR(\tilde{B}\to Zb) /
BR(\tilde{B}\to Hb) \approx 2/1/1. \label{2:1:1}
\end{equation}
The large coupling between $\tilde{B}$ and $B^+$ forces a sizeable
mixing between the two and the lightest mode then inherits part of the
peculiar (because of the large branching ratio) decay mode of $B^+$
into $H b$. This effect is suppressed by powers of
$M_{\mathcal{Q}^\prime}$ but the large value of the couplings and the
fact that it is competing with the suppressed coupling of $\tilde{B}$
with $b$, means that very large values of $M_{\mathcal{Q}^\prime}$ are required
before the decoupling is effective. 
In order to be more specific, let us focus on the physical Yukawa
couplings (after EWSB)
between the bottom quark and the lightest new quark of charge $-1/3$,
that we will call $B_l$. The relevant part of the Lagrangian reads
\begin{equation}
\mathcal{L}= \frac{H}{\sqrt{2}} \bar{b} \left[ 
\lambda_{bB_l} P_R  
+\lambda_{B_l b} P_L  
\right] B_l+\mathrm{h.c.}+\ldots,
\end{equation}
where $P_{L,R}=(1\mp \gamma^5)/2$ are the standard chirality projectors.
In the slow decoupling limit of Eq.(\ref{slow:decoupling}) we can
obtain approximate analytic expressions for these couplings
\begin{eqnarray}
\lambda_{bB_l}&=& - Y_B s_2 c_{bR} 
+ \frac{1}{2}Y_B^3 s_2 c_{bR}(1-3s_{bR}^2)
\left(\frac{s_2^2}{2}\frac{v^2}{M_{\tilde{B}}^2}-
\frac{v^2}{M_{\mathcal{Q}^\prime}^2}\right)
\nonumber \\
&&
- Y_B^3 s_2 (2-3s_{bR}^2) \frac{v^2}{M_{\tilde{B}}M_{\mathcal{Q}^\prime}}
+\mathcal{O}\left(\frac{v^4}{M^4},\frac{v^2M_{\tilde{B}}}{M_{\mathcal{Q}^\prime}^3}
\right), \label{lambdabB}\\
\lambda_{B_l b}&=& - Y_B^2 s_2^2 s_{bR}
c_{bR} \frac{v}{\sqrt{2}M_{\tilde{B}}}
+2\sqrt{2} Y_B^2 s_{bR} \frac{v}{M_{\mathcal{Q}^\prime}}
+\mathcal{O}\left(\frac{v^3}{M^3},\frac{v M_{\tilde{B}}}{M_{\mathcal{Q}^\prime}^2}
\right).   \label{lambdaBb}
\end{eqnarray} 
The first term in Eq. (\ref{lambdabB}) is the one that would determine
the decay pattern of $B_l$ if it came mainly from a vector-like
singlet, resulting in the well known $2:1:1$ pattern, see
Eq. (\ref{2:1:1}). This coupling receives corrections from the
presence of $B^+$, like the one in the second line of
Eq. (\ref{lambdabB}) that can easily exceed the otherwise leading
term. Even more important in our example is the fact that the
$\lambda_{B_lb}$, which is irrelevant in the case of a vector-like
singlet, receives huge corrections that can dramatically change the
decay pattern of $B_l$. 
\begin{figure}[ht]
{\includegraphics[width=0.45\columnwidth]{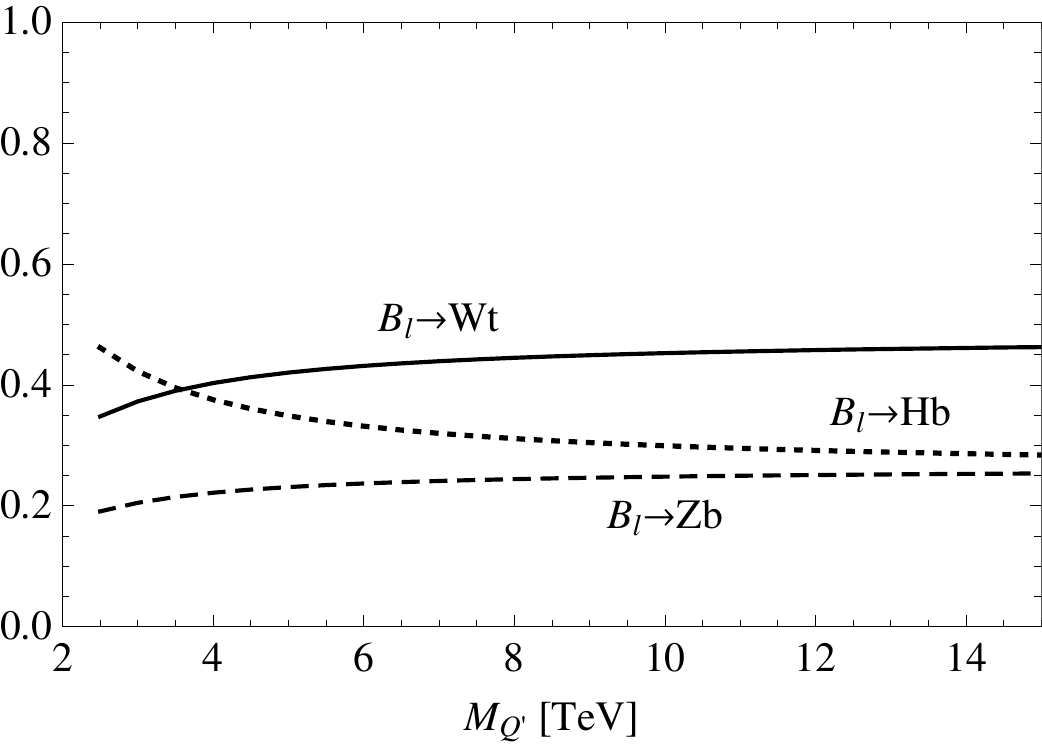}
\hfil
\includegraphics[width=0.45\columnwidth]{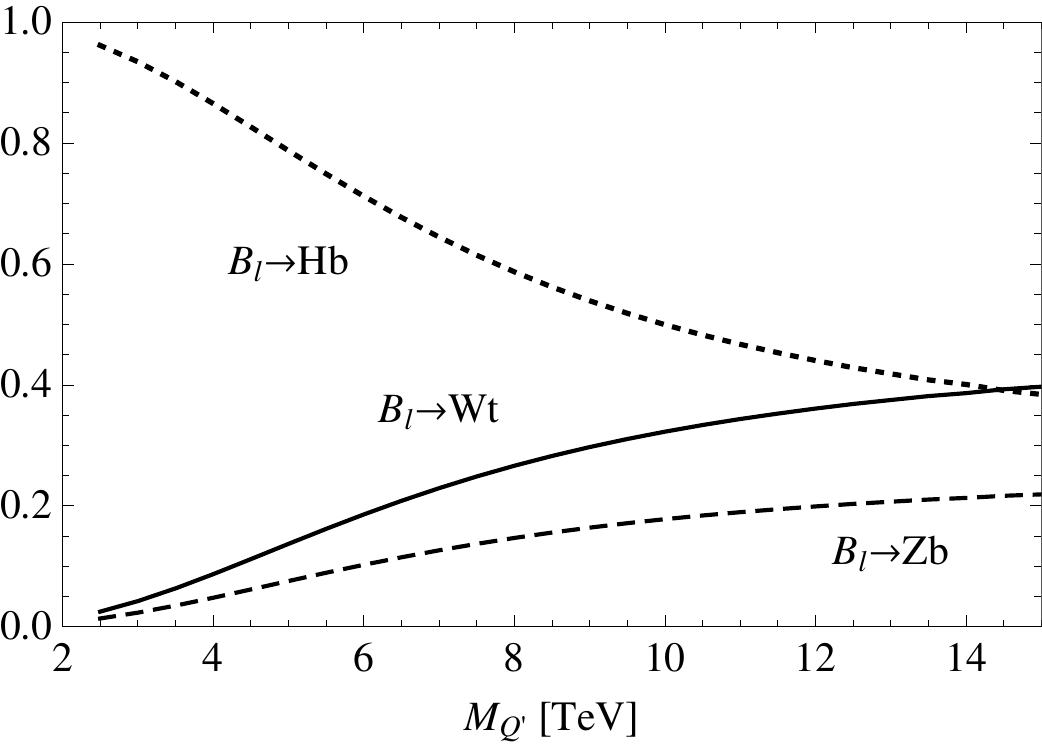}
}
\caption{Branching fraction decay of $B_l$ as a function of
$M_{\mathcal{Q}^\prime}$ for $s_{bR}=0.05$ (left) and $s_{bR}=0.2$ (right). We have
fixed $M_{B_l}=1$ TeV and $M_{\mathcal{Q}}=M_{\mathcal{Q}^\prime}$.
}\label{slowdecoupling:fig}
\end{figure}
Explicitly, for the leading correction, which
corresponds to the second term in Eq. (\ref{lambdaBb}) to be much smaller
than the first term in $\lambda_{bB_l}$, so that the decay pattern
becomes again the standard one, we need the mass of $B^+$ to be
\begin{equation}
M_{\mathcal{Q}^\prime} \gg \frac{2 s_{bR}^2}{c_{bR}} \frac{Y_B^2
v^2}{m_b}
\approx 272 \, \frac{
s_{bR}^2}{c_{bR}} \left(\frac{Y_B}{3}\right)^2 \mbox{ TeV},
\end{equation}
where we have used that the bottom quark mass is approximately given by
\begin{equation}
m_b \approx Y_B s_2 s_{bR} \frac{v}{\sqrt{2}}.
\end{equation}
This is a very conservative estimate of the mass scale at which $B^+$
stops having a profound impact on the decay pattern of $B_l$. A more
quantitative result is given in Fig.~\ref{slowdecoupling:fig} in which
we show the branching ratios of $B_l$ as a function of
$M_{\mathcal{Q}^\prime}$ for $s_{bR}=0.05$ (left) and $s_{bR}=0.2$
(right). We
have chosen the value of 
$M_{\tilde{B}}$ so that the mass of the lightest new quark
is $M_{B_l}=1$ TeV. All the remaining masses are set equal to
$M_{\mathcal{Q}^\prime}$. As we see, even for very small values of
$s_{bR}\sim 0.05$ extra quarks with masses in the $3-5$ TeV region still
have an important impact on the decay pattern of $B_l$. This moves up
to $15$ TeV for $s_{bR}=0.2$. In fact, for $s_{bR}\gtrsim 0.2$ 
we could have the challenging situation in
which the only discovered new particle at the LHC is $B_l$ but its
decay patterns differ dramatically from the ones of a vector-like
singlet. This is due to its mixing with new quarks with masses
 $\sim 10-20$ TeV and would therefore escape experimental scrutiny
even with an upgraded energy phase of the LHC.

\section{Singlet scalar searches in Composite Higgs Models \label{appendix:B}}

Non-minimal composite Higgs
models~\cite{Gripaios:2009pe,Mrazek:2011iu,Frigerio:2012uc,Bertuzzo:2012ya,Chala:2012af,Vecchi:2013bja}
can contain extra neutral 
singlets $\eta$ in the spectrum of pseudo-Nambu-Goldstone bosons. This
happens for instance in the case of the
$SO(6)/SO(5)$~\cite{Gripaios:2009pe} or
$SO(7)/G2$~\cite{Chala:2012af} cosets, in which the scalar Lagrangian
has an $\eta\rightarrow -\eta$ symmetry that is only broken by couplings to
the SM elementary fermions\footnote{A vacuum expectation value for
$\eta$ could generate an $\eta H H$ coupling from the loop-induced
scalar potential. In explicit models, however, a large region of parameter
space is compatible with $\langle \eta\rangle =
0$~\cite{Redi:2012ha}}. Thus, $\eta$ can
couple linearly to fermions but with a coupling suppressed by a
factor $1/f$, with $f$ the compositeness scale, through the operator
\begin{equation}
\mathcal{O} = \frac{c
Y}{f}\eta \bar{\psi}_L H \psi_R,
\end{equation}
where $Y$ is the fermion Yukawa coupling and $c$ is expected to be
order one (other dimension 5 operators are equivalent to this one via the
classical equations of motion).
The main standard production mechanism for $\eta$ would then be gluon
fusion but with a rate that is suppressed by a factor $v^2/f^2$ with
respect to the SM Higgs gluon fusion production. In
addition, the main decay of $\eta$ is into a $b\bar{b}$ final state
for masses below $\sim 350$ GeV, which suffers from a huge QCD
background. 
Thus, in these cases, process similar to the one we have
considered in this work, with the replacement of $H$ with $\eta$,
\begin{equation}
pp\to G^*\to B_\eta \bar{b}+\bar{B}_\eta b
 \to \eta b \bar{b} \to  4b,
\end{equation} 
could provide the leading channel to discover the composite singlets.
\newpage

\bibliography{mybib}

\end{document}